\documentclass[prb,twocolumn,aps,superscriptaddress,showpacs]{revtex4}
\usepackage{times,xspace}
\usepackage{graphicx,graphics,color,epsfig}
\usepackage{amsmath}
\usepackage{amssymb}
\usepackage{amsfonts}
\usepackage{amsfonts}
\usepackage{epstopdf}
\usepackage{bm}
\usepackage{color}
\usepackage{float}

\def\be{\begin{equation}}
\def\ee{\end{equation}}
\def\ba{\begin{eqnarray}}
\def\ea{\end{eqnarray}}

\newcommand{\bk}{\mathbf{k}}
\newcommand{\bp}{\mathbf{p}}

\begin{document}
\title{Quantum anomalous Hall effect with field-tunable Chern number near Z$_2$ topological critical point}

\author{Le Quy Duong}
\affiliation{Centre for Advanced 2D Materials and Graphene Research Centre,
National University of Singapore, Singapore 117546}
\affiliation{Department of Physics, National University of Singapore, Singapore 117542}

\author{Hsin Lin}
\affiliation{Centre for Advanced 2D Materials and Graphene Research Centre,
National University of Singapore, Singapore 117546}
\affiliation{Department of Physics, National University of Singapore, Singapore 117542}

\author{Wei-Feng Tsai}
\email{wftsai@mail.nsysu.edu.tw}
\affiliation{Department of Physics, National Sun Yat-sen University, Kaohsiung 80424, Taiwan}

\author{Y. P. Feng}
\affiliation{Centre for Advanced 2D Materials and Graphene Research Centre, National University of Singapore, Singapore 117546}
\affiliation{Department of Physics, National University of Singapore, Singapore 117542}

\date{\today}
\begin{abstract}
We study the practicability of achieving quantum anomalous Hall (QAH) effect with field-tunable Chern number in a magnetically doped, topologically trivial insulating thin film. Specifically in a candidate material, TlBi(S$_{1-\delta}$Se$_{\delta}$)$_2$, we demonstrate that the QAH phases with different Chern numbers can be achieved by means of tuning the exchange field strength or the sample thickness near the Z$_2$ topological critical point. Our physics scenario successfully reduces the necessary exchange coupling strength for a targeted Chern number. This QAH mechanism differs from the traditional QAH picture with a magnetic topological insulating thin film, where the ``surface'' states must involve and sometimes complicate the realization issue. Furthermore, we find that a given Chern number can also be tuned by a perpendicular electric field, which naturally occurs when a substrate is present.
\end{abstract}

\maketitle
\section{Introduction}
Topological insulators (TIs) belong to a new phase of matter with their topological nature protected by time reversal symmetry (TRS)\cite{HasanRMP,ShouChernZhangRMP,AndoJPSJ}. Usually, spin-orbit coupling (SOC) plays a key role to ``twist'' the Bloch wavefunctions in the first Brillouin zone of the solids and such a twist can be characterized by a Z$_2$ number. Materials with a non-trivial Z$_2$ number, such as Bi$_2$X$_3$ (X = Se, Te) \cite{TInature,Xianature,YLChen,HsiehPRL,Jeffreypaper}, have attracted strong interest in recent years due to their unusual spin-momentum locked, massless Dirac-like boundary states with potential for spintronic applications \cite{MacDonaldreview,NikolicPRL,DahmPRApp} and realizing other exotic phenomena originally appeared in high energy physics \cite{majorana,Topologicalfieldtheory,monopole,axion,Witteneffect,Oshikawa}. 

Moreover, the TIs host variety of other topological phases. One of the significant examples with broken TRS is the quantum anomalous Hall (QAH) effect\cite{FuKane,RuiYu}. This effect is first proposed by Haldane in a two-dimensional (2D) honeycomb lattice with locally non-vanishing magnetic flux but zero in average\cite{Haldane88}. Similar to the integer quantum Hall effect, it can exhibit the quantized Hall conductance, $C\frac{e^2}{h}$, through gapless edge channels with an integer number $C$ (the so-called Chern number\cite{Kubo}) indicating its non-trivial bulk topological property without Landau levels. The existence of the edge channels makes it even more attractive due to its predicted applications for dissipationless electronics and integrated circuits \cite{RuiYu,CZChang}. So far, without using TIs, several approaches have been proposed to realize QAH state, for instance, via mainly band structures like in the mercury-based quantum wells with ferromagnetic (FM) order \cite{HgWQ} and in graphene with Rashba interactions and the exchange field \cite{grapheneQAH}; via electron-electron interactions like in transition metal oxide heterostructures \cite{interfaceengineering}, nearly flat band systems\cite{flatband}, and quadratic band-crossing systems\cite{quadraticbandA,quadraticbandB}; via disorder effect in Anderson insulator\cite{disorder}. However, they have not yet been measured by experiments. 

By contrast, the QAH phases can also be realized by using the TIs as a platform. Theoretically, the QAH effect can be achieved either by gapping out the surface states of the three dimensional (3D) TIs\cite{RuiYu} or by inserting ferromagnetically ordered dopants in the TI thin films \cite{CZChang,trajectory,scaleinvariant}, referring to the same mechanism: the spin-polarized band inversion from the interplay between the FM ordering and the SOC in the TIs. Recently, breakthrough experiments have realized the QAH phase with $C=1$ in a magnetic TI thin film of Cr-doped (Bi,Sb)$_2$Te$_3$ \cite{CZChang,trajectory,scaleinvariant} and with a large Chern number in {\it bosonic}, photonic crystals\cite{Photoniccrystal}. Hence, these significant results are motivation to disclose these following questions for {\it fermionic} systems: 1) How to obtain a QAH phase with higher Chern number? 2) Is it possible to tune the Chern number by other perturbations? 3) Is measuring the magnetic TI thin films a key component to obtain QAH effect?
   
In this paper, we provide an alternative route, compared to the previous studies \cite{Niu12,Jwang329,HighChern2}, to answer the aforementioned questions. We show the feasibility of exhibiting QAH effect with field-tunable Chern number in a magnetically doped, topologically {\it trivial} insulating thin film, in proximity to the Z$_2$ topological critical point \cite{Pallabprl}. Specifically, we will focus on a candidate compound, TlBi(S$_{1-\delta}$Se$_{\delta}$)$_2$, for its tunable property from a trivial insulator to an non-trivial one by changing the chemical composition \cite{sciencepaper}. Although the microscopic mechanism to realize the QAH phases is still the same as the magnetically doped TIs, our scenario further offers a few attractive features to achieve higher Chern number: (1) Making the system close to the Z$_2$ critical point lowers the threshold value of the exchange field to induce the high order spin-polarized band inversion;
(2) we start with a {\it trivially insulating} thin film, in sharp contrast to previous studies. As a result, all 2D-like subbands have bulk nature and it has advantages to avoid the treatment for the sample surfaces, usually exposed to complex environment; (3) we also suggest an implementation of the potential gradient or the dielectric substrate to reduce the critical value of the exchange field. This has not yet been explored in previous related studies. Therefore, in brief, our work strongly suggests that the trivial insulating thin film with sufficiently large SOC can provide a unique platform to exhibit QAH effect with high Chern number.

The remainder of this paper is organized as follows. In section II, we introduce our effective model for the thin films and the methods for demonstrating the topological properties of the system. In section III, in the absence of any exchange/electric field we discuss electronic properties of the effective model with an emphasis on its tunable topological phase transition by the inter-layer tunneling parameter. In section IV, we present our scenario to achieve the QAH effect via tuning either an exchange field or a perpendicular electric field. Lastly, we give a brief conclusion in section V.

\section{Model and Method}
Many of the usual 3D TI materials are layered materials. For example,  Bi$_2$Se$_3$ can be viewed as consisting of Se-Bi-Se-Bi-Se quintuple layers stacked along [111] ($z$) direction.\cite{TInature,Xianature} This picture is also applicable for our focused, Tl-based compounds TlBi(S$_{1-\delta}$Se$_{\delta}$)$_2$, in which the layer unit becomes Tl-(Se,S)-Bi-(Se,S).\cite{sciencepaper,BiTlSSe} When they are confined along $z$ direction to be thin films, the underlying mechanism for the ground state to becoming topologically non-trivial could be more insightful if one considers each one-unit-cell layer as a building block and then couples them with nearest-neighbor inter-layer hopping \cite{DasTI,LiangOdd}. As a representative sample to achieve the QAH effect, we start with an effective one-unit-cell layer $\bk\cdot\bp$ Hamiltonian $H_{l}$ around $\Gamma$ point,
\begin{equation}
H_{l}= \begin{pmatrix} 
\frac{k^2}{2m}     & vk_{+}            & D_k            &  0 \\
vk_{-}             & \frac{k^2}{2m}    & 0              & D_k \\ 
D_k                & 0                 & \frac{k^2}{2m} & -vk_{+} \\
0                  & D_k               & -vk_{-}        &\frac{k^2}{2m} 
\end{pmatrix},
\end{equation}
where the basis vector for $l$th layer is taken to be $\phi^\dagger_{l\bk}=[a^\dagger_{l\bk\uparrow},a^\dagger_{l\bk\downarrow},
b^\dagger_{l\bk\uparrow},b^\dagger_{l\bk\downarrow}]$ with $a$ and $b$ operators corresponding to local $p_z$ orbitals of Se or S residing on the top and bottom  unit layer hybridized with neighboring atomic orbitals within it. $\bk$ denotes a 2D momentum with magnitude $k=\sqrt{k_x^2+k_y^2}$ and $k_{\pm}=k_x \pm i k_y$; $m$ represents the fermion effective mass and $v$ is determined by the SOC. The presence of $D_k=d+\frac{k^2}{m_d}$ in $H_l$ is due to the non-vanishing overlap between the top and the bottom orbitals within the layer. The structure of $H_l$ can be clearly understood as two coupled Rashba systems (with opposite chiralities), which preserves both TRS and inversion symmetry (IS), forming a key component in TI thin films.

\begin{figure}[t]
\rotatebox[origin=c]{0}{\includegraphics[width=.98\columnwidth]{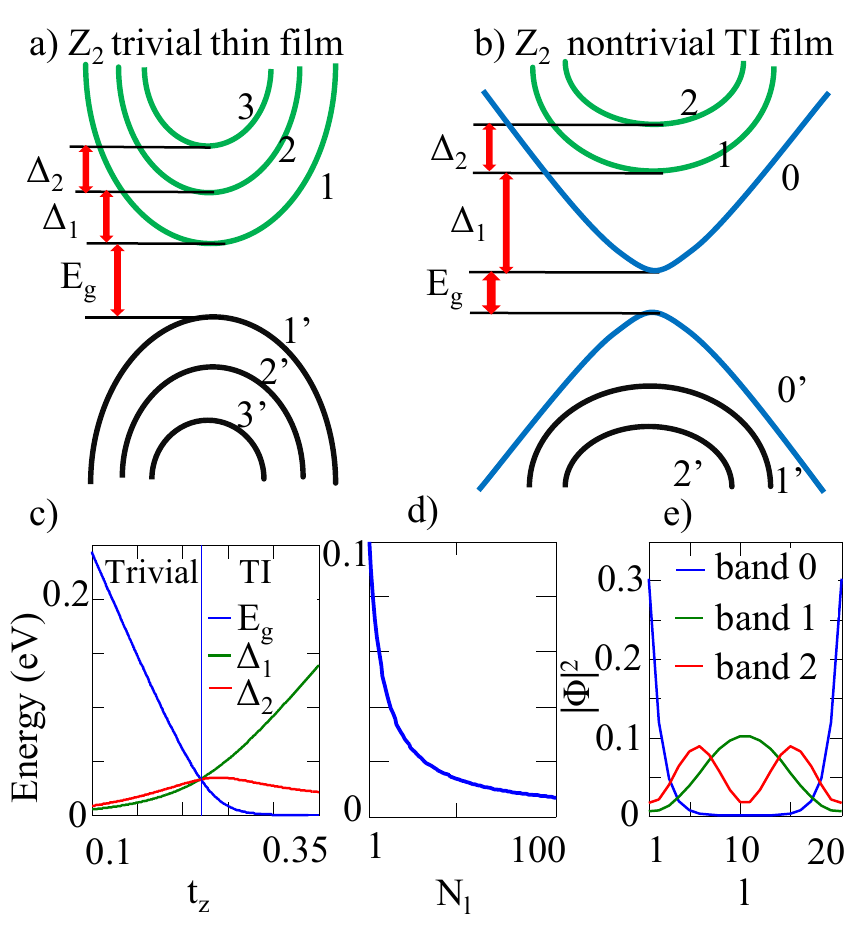}}
\caption{(Color online) 
The schematic energy spectra of the thin film model around $\Gamma$ point from a trivial insulator (a) to a Z$_2$ TI (b) with two types of energy scales: the system band gap $E_g$ and the energy difference between neighboring bands $\Delta_i$. The positive integer numbers without (with) prime denote conduction (valence) bands. ``$0$'' and ``$0^\prime$'' denote bands with surface nature. (c) The calculated $E_g$ and $\Delta_i$ as a function of $t_z$. $t_{zc}$ is found to be 0.22 eV here. (d) The band gap $E_g$ of the thin film at $t_z=t_{zc}$ as a function of the number of layers $N_l$. (e) The wave function distribution $|\Phi|^2$ at $\Gamma$ point as a function of $l$ for the lowest three conduction bands in an $N_l=20$, Z$_2$ TI thin film with $t_z=0.23$ eV ($l$ is the layer index). Other numerical parameters are given in the main text.} 
\end{figure}
To obtain the full Hamiltonian $H$ for the thin films, we then stack each quintuple layer along $z$ direction by coupling nearest-neighbor layers with a finite, spin-non-flip tunneling term $T$ as follows
\begin{equation}
H=\begin{pmatrix}
H_{1}       & T           & 0     & \cdots \\
T^{\dagger} & H_{2}       & T     & \cdots \\
0           & T^{\dagger} & H_{3} & \cdots \\
\vdots      & \vdots      &\vdots & \ddots \\
\end{pmatrix},
\label{eq2}
\end{equation}
where
\begin{equation}
T= \begin{pmatrix} 
0            & 0            & 0  & 0 \\
0            & 0            & 0  & 0 \\ 
t_{z}        & 0            & 0  & 0 \\ 
0            & t_{z}        & 0  & 0 
\end{pmatrix}.
\end{equation}   
Note that the tunneling between top and bottom positions of the two adjacent layers is $t_z$. The real-space coordinates in $z$ direction are positive integers with each number corresponding to $l$th layer. This model is seen to correctly give the spin texture for the surface Dirac cone states in Bi$_2$Se$_3$ and has been applied to transport simulations of a TI slab.\cite{GauravPRB} When taking into account the Zeeman effect introduced by the ferromagnetic order (with polarization along $z$ direction), as we shall consider later on, we further add in an extra term, $H_M=M_z\sigma_z$ for $H_l$ with $\sigma_z$ acting on the spin space. 

The parameter specific model for TlBi(S$_{1-\delta}$Se$_{\delta}$)$_2$ in the trivial side is close to the $Z_2$ phase transition critical points. We have $m=0.0625$ eV$^{-1}$\AA$^{-2}$; $m_{d}=-0.04$ {eV}$^{-1}${\AA}$^{-2}$; $d=-0.22$ eV, $v=2.5$ eV$\cdot${\AA}; $t_{z}=0.2$ eV.\cite{preform} By increasing $t_z$, we can obtain a nontrivial $Z_2$ topological insulator phase. This model has been used to describe the topological phase transition in TlBi(S$_{1-\delta}$Se$_{\delta}$)$_2$ and delivered the spin-polarized surface related states both in the trivial and the nontrivial region in good agreement with recent spin- and angle-resolved photoemission measurements.\cite{preform} The Hall conductivity in unit of $\frac{e^2}{h}$ is calculated based on Kubo formalism\cite{Kubo,titus} which is an integration of the Berry phase of Bloch wave function. The Hall conductivity is thus:
\begin{multline}
\sigma_{xy}=e^2\hbar\sum_{m,n,\bk}\frac{\text{Im}[\langle n\bk|V_{x}|m\bk\rangle\langle m\bk|V_{y}|n\bk\rangle]}{(E_{n\bk}-E_{m\bk})^{2}}.\
\\(n_{f}(E_{n\bk})-n_{f}(E_{m\bk})),
\end{multline}
where $m,n$ are band indices, $V_{x,y}$ are velocity operators, and $n_{f}(E_{n\bk})$ stands for Fermi-Dirac distribution function at eigen-energy $E_{n\bk}$.    

\section{Topological phase transition: From trivial to non-trivial insulators}
It would be helpful to begin with a warm-up example to demonstrate how the effective model can be driven into non-trivial Z$_2$ insulators by tuning the parameter $t_z$ which is related to the lattice constants and the size of atomic orbitals in general. For concreteness, we adapt the parameters mentioned above which are fitted to band structures of BiTl(S$_{1-\delta}$Se$_{\delta}$)$_2$\cite{BiTlSSe,preform}. However, it is worth mentioning that this model is also applicable for the other Bi$_2$X$_3$ compounds. \cite{GauravPRB}

The typical energy spectra around $\Gamma$ point for Z$_2$ trivial and non-trivial thin films are shown in Figs. 1(a) and 1(b), respectively. Without breaking both TRS and IS, each band in the spectra is obviously spin degenerated and thus has a zero expectation value for the net spin polarization. By defining $E_g$ as the energy gap between the lowest conduction band and the highest valence band at $\Gamma$ point in the thin film, and $\Delta_i$ as the gap between $i$th lowest conduction band and $i+1$th lowest one, on the trivial insulator side, we find that $E_g$ is typically larger than $\Delta_i$, while among $\Delta_i$ they are comparable, as a result of all quantum well states.

The aforementioned feature can be changed via chemical (Se) doping, or effectively, increase the tunnelling energy $t_z$. Close to a ``critical'' value of $t_z$ (called $t_{zc}$, at which the 3D {\it bulk} band gap closes, independent of the thickness of the thin films), $E_g$ becomes comparable to $\Delta_1$ and turns into a smaller value rapidly after passing $t_{zc}$, as can be seen in Figs. 1(b) and 1(c). This crossover phenomenon between the values of $E_g$ and $\Delta_1$ marks a topological phase transition (TPT): From a thin film of trivial insulator to a thin film of a Z$_2$ topological insulator. The latter phase is sharply identified by the presence of the surface states, a consequence of the band inversion, with Dirac-like energy dispersion [see the band labeled by ``$0$'' and ``$0^\prime$'' in Fig. 1(b)]. Note that the surface states are distinct from the usual quantum well (bulk) states, because their wave functions would be almost localized at either the top or the bottom layer, as compared in Fig. 1(e) by showing wave function distribution $|\Phi|^2$ as a function of the layer index $l$. 


There are a few things in the TI regime worth mentioning here. Firstly, when $t_z > t_{zc}$, it is understood that the massiveness of the Dirac spectrum for the surface states is due to the tunneling barrier (or inevitable wave function overlapping between boundaries), that is, with a relatively small $t_z$ (but still $>t_{zc}$), in thin films. Secondly, similar to the usual case for growing Bi$_2$Se$_3$ thin films\cite{crossover}, the Dirac mass can be reduced to zero by increasing the number of layers $N_l$ to a $t_z$-dependent threshold value $N_{lc}$. For instance, as $t_z$ approaches to $t_{zc}$ from above, the energy gap is closed only when $N_{lc}$ goes to infinity in the true 3D limit [see Fig. 1(d)]. Finally, it is quite important to notice that typically $E_g < \Delta_2 < \Delta_1$ for TIs, while near $t_z\sim t_{zc}$ even on the trivial insulating side, these energy scales are all comparable with each other. This is the key observation of our proposed scenario for realizing QAH effect with field-tunable Chern number, as we will explain below. 
   
\begin{figure}[t]
\rotatebox[origin=c]{0}{\includegraphics[width=.98\columnwidth]{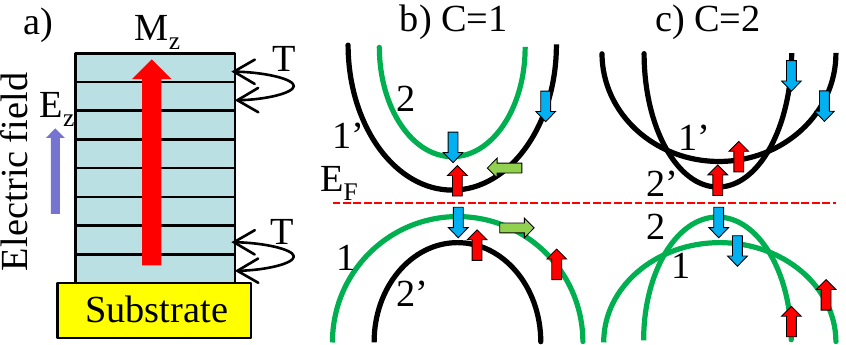}}
\caption{(Color online)
The left panel (a) schematically shows our proposed scenario to achieve the field-tunable QAH effect: Let layered TlBi(S$_{1-\delta}$Se$_\delta$)$_2$ thin film with ferromagnetic order be grown on a substrate, which provides an electric potential along $z$ direction. The right panel, (b) and (c), shows schematic band structures, spin-texture (``skyrmion''-like), and band-inversion with suitable band labels for $C=1$ and $C=2$ QAH phases, respectively, at $t_z<t_{zc}$.
} 
\end{figure}

\section{QAH effect from trivial insulating thin films}
In this section, we demonstrate how the QAHE with field-tunable Chern number can be achieved theoretically through our scenario. As shown schematically in Fig. 2(a), the TlBi(S$_{1-\delta}$Se$_{\delta}$)$_2$ thin film, growing on a substrate, on the {\it trivial} insulating side with $t_z\sim t_{zc}$ (doping-tunable) is taken to be our prototype sample. During the process toward QAH effect, we also assume that certain magnetic dopant can be distributed homogeneously over the whole sample and are ferromagnetically ordered to provide the necessary exchange field. In the presence of the exchange field, the core concept to exhibit QAH effect is simply from the band inversion phenomenon occurred between pairs of the conduction and valence bands with different spin polarizations. Figs. 2(b) and 2(c) schematically show one pair of the bands inverted (thus with Chern number $C=1$) and two pairs inverted (with Chern number $C=2$), respectively. After band inversion, each band forms a skyrmion-like spin texture around $\Gamma$ point in momentum space, leading to non-vanishing Chern number. In addition, in order to achieve the QAH phase with even higher Chern number, more pairs of the band inversions are needed. According to the comparison among various energy scales in the previous section, with a given strength of the exchange field $M_z$, our focus on the trivial side near $t_{zc}$ would more likely arrive at a phase with high $C$.  

\subsection{Exchange field tunable QAH effect}
We now show the band structure evolution around $\Gamma$ point as a function of $M_z$  for a TlBi(S$_{1-\delta}$Se$_{\delta}$)$_2$ thin film with $N_l=6$ in Fig. 3.\cite{warping} Fermi level $E_F$ is always set at zero energy and the labels $i=1,2$ ($i=1^\prime,2^\prime$) denote the $i$ th lowest (highest) conduction (valence) bands, in the {\it absence} of $M_z$, with spin-up and spin-down shown in different colors.

Starting with $M_z=0$, each band is spin-degenerate and the system is in the $C=0$ (trivial) phase. Increasing $M_z$ causes spin splitting, which shifts the bands with spin-up and spin-down polarizations in opposite directions with respect to $E_F$, and hence reduces $E_g$ [Fig. 3(b)]. As $M_z$ reaches $M_{c1}$, the energy gap is closing with vanishing out-of-plane spin moment at the touching point [Fig. 3(c)]. Further increasing $M_z$ reopens the gap again, forms skyrmion-like spin texture, and indicates a topological phase transition due to band inversion from $C=0$ trivial phase to $C=1$ QAH effect [Fig. 3(d)]. We note that now the band labeled ``1'' and ``$1^\prime$'' are switched. As $M_z=M_{x1}$, the band ``2'' (``$2^\prime$'') meets with ``$1^\prime$'' (``1'') [Fig. 3(e)] and the band ``2'' with spin-down and ``$2^\prime$'' with spin-up become prominent near $E_F$ if $M_z$ continues increasing until $M_z=M_{c2}$ [Fig. 3(f)]. At $M_z=M_{c2}$, the gap is closing [Fig. 3(g)] and implies a second TPT, which adds the Chern number by one and hence $C=2$ after the gap-reopening as $M_z>M_{c2}$ [Fig. 3(h)]. The fashion shown here is in fact quite similar to some previous proposals for getting the QAH effect with high Chern number in the TI thin film regime \cite{Niu12,Jwang329,HighChern2}.

\begin{figure}[t]
\rotatebox[origin=c]{0}{\includegraphics[width=.95\columnwidth]{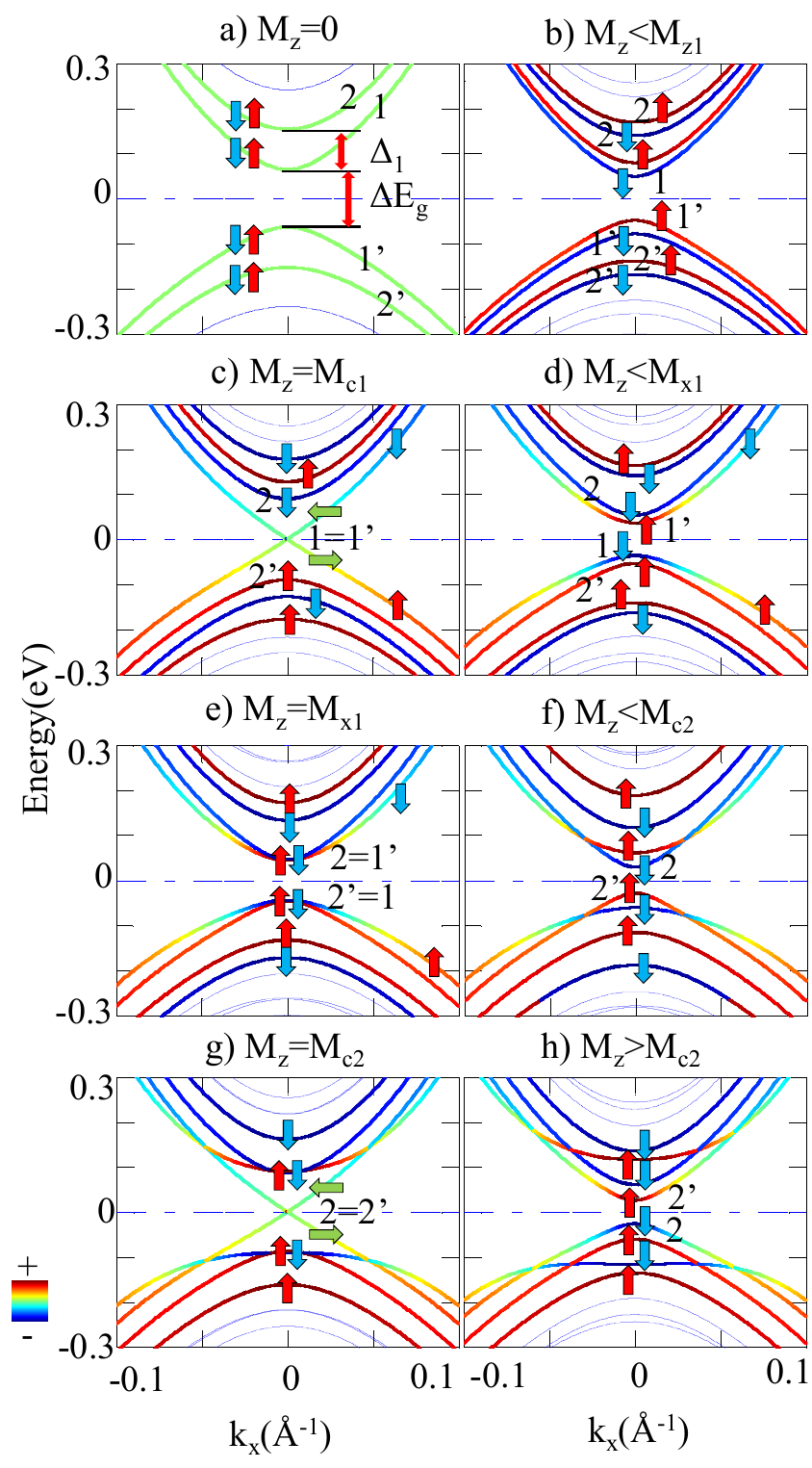}}
\caption{(Color online)
Evolution of the sub-band structure at $t_z=0.2$ eV and $N_l=6$ around $\Gamma$ point upon increasing exchange field $M_z$. The band labels are used as usual. Additionally, the color dressing in each band represents the momentum-dependent, out-of-the-plane spin polarization with arrows indicating the main spin-polarized component. (a) $M_{z}=0$; (b) $0 < M_{z} < M_{c1}$ before first topological phase transition (TPT); (c) $M_z=M_{c1}$, where the gap closes at the first time; (d) $M_{c1} < M_{z} < M_{x1}$ in the QAH phase with $C=1$; (e) $M_{z}=M_{x1}$, where two lowest conduction bands meet; (f) $M_{x1} < M_{z} <M_{c2}$ and the system approaches second band inversion; (g) $M_{z}=M_{c2}$ at another critical point; (f) $M_{z} > M_{c2}$, where the second TPT occurs, and the system enters QAH state with $C=2$.} 
\end{figure}

\subsection{Toward higher Chern number}
The model simulation and discussion in the previous subsection give us a clear physical picture of our mechanism to obtain high Chern number. Given sufficiently large exchange field, the band gap can close and reopen multiple times due to the presence of relatively intensive 2D subbands (quantum well states), which is a consequence of a trivial insulating sample with (nearly) critical Se-doping. 

This mechanism is completely based on the ``twist'' of the spin-polarized {\it bulk} states in the quasi-2D system. Thus, it is distinctive from the original proposal, where the QAH effect is achieved by gapping out the surface Dirac cones in 3D TIs \cite{FuKane,DHLee}. To see this, we first present energy spectrum around $\Gamma$ point and the corresponding spin texture for $C=2$ QAH state with $N_l=20$, {\it i.e.}, in the thick film limit [see Fig. 4(a)]. The initially lowest, spin-down polarized conduction band now becomes the third highest valence band, indicating that the spin-polarized bands inverted twice. To explain the underlying physics, from the wave function distribution as a function of $l$ in Fig. 4(b), the first three lowest conduction bands all have bulk nature. Similar properties are also found for the other bands. Such a feature results in a subtle but important difference from previous studies\cite{Niu12,Jwang329}, because in our case no surface bands are involved in the whole process. As we will consider later, this might affect the real experiments in which each sample is usually grown on certain substrate. 

\begin{figure}[t]
\rotatebox[origin=c]{0}{\includegraphics[width=.98\columnwidth]{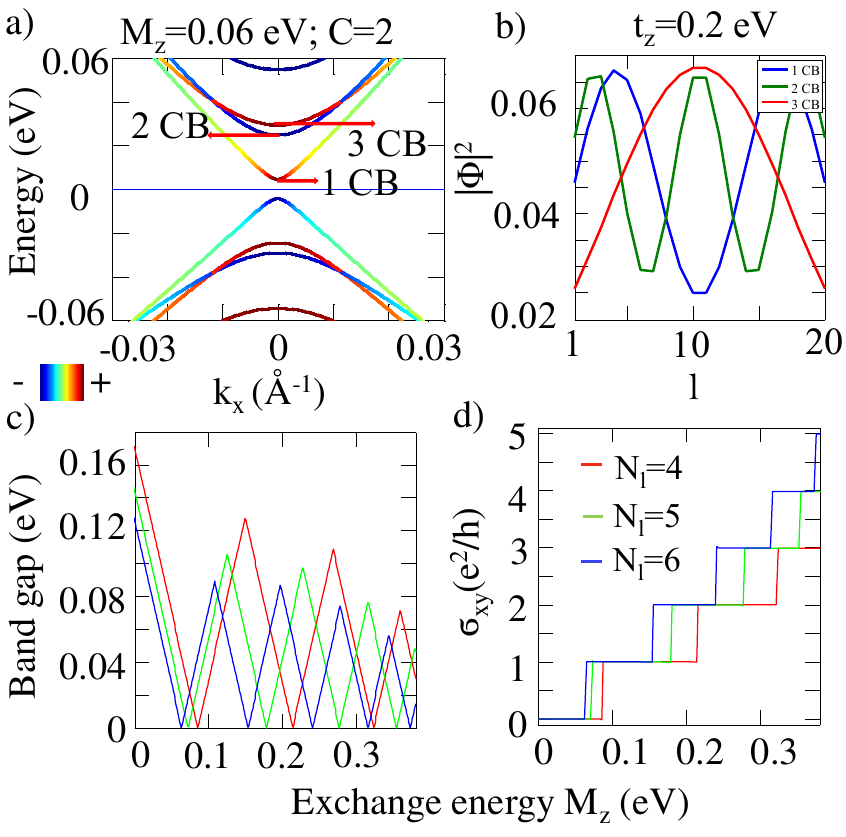}}
\caption{(Color online) (a) Band structure of the QAH phase with $C=2$ in the thick film limit, $N_l=20$ ($M_z=0.06$ eV) and the color dressing is used for spin polarizations. (b) The corresponding wave function distributions of the three lowest conduction bands at the $\Gamma$ point as a function of layer index $l$. (c) and (d) show the band gap evolution and the quantized Hall conductivity, respectively, as a function of $M_z$ for the given $N_l=4,5,6$ cases. All plots here use $t_z=0.2$ eV.} 
\end{figure}

To invert more spin-polarized bulk bands, our mechanism suggests at least two ways: 1) Apply large exchange field on the sample, and 2) increase the thickness of the sample. In Figs. 4(c) and 4(d), we explicitly calculate the band gap and the Hall conductivity as a function of $M_z$, respectively. The gap closes and reopens multiple times with the presence of the corresponding quantized plateaus in $\sigma_{xy}$, indicating a rich phase diagram of the QAH system. The Chern number increases in one integer step when $M_z$ increases, a similar trend compared with the usual quantum Hall system. In addition, in the same figures by using different colors we also present both quantities with different number of layers. Clearly, for a thicker film with a given $M_z$, it is more likely to end up with a higher Chern number insulator due to the shrinking of $\Delta_i$, which is inversely proportional to $N_l$.

From recent experiments in magnetic topological insulators such as Cr or Fe doped (Bi, Sb)$_2$Te$_3$, people observed that these magnetic dopants can be ferromagnetically ordered at temperature of order 100 K.\cite{Svanda01,Zhou06,Chang13,Kou14} The corresponding effective exchange field strength $M_z$ can be estimated as large as 0.2 eV with 10\% doping\cite{RuiYu,Jwang329} and thus strongly indicates the feasibility of our scenario. To roughly estimate what the largest Chern number could be achieved, one can simply count how many quantum well states (subbands), labeled $x$, are able to be inverted by applying $M_z$. By noticing that $E_g\sim\Delta_i\sim 0.035$ eV in Fig. 1(c) with $N_l = 20$, the number can be estimated through the following formula, $x = [(M_z -0.0175)/0.035 +1]$, where [$\cdots$] denotes a floor function. Inserting the value of $M_z \approx 0.2$ into the formula yields $x = 6$, {\it i.e.}, the largest $C = 6$ in this case. In addition, the energy range for this QAH phase to be stable could be as large as 0.035 eV, above room temperature.

\begin{figure}[t]
\rotatebox[origin=c]{0}{\includegraphics[width=.98\columnwidth]{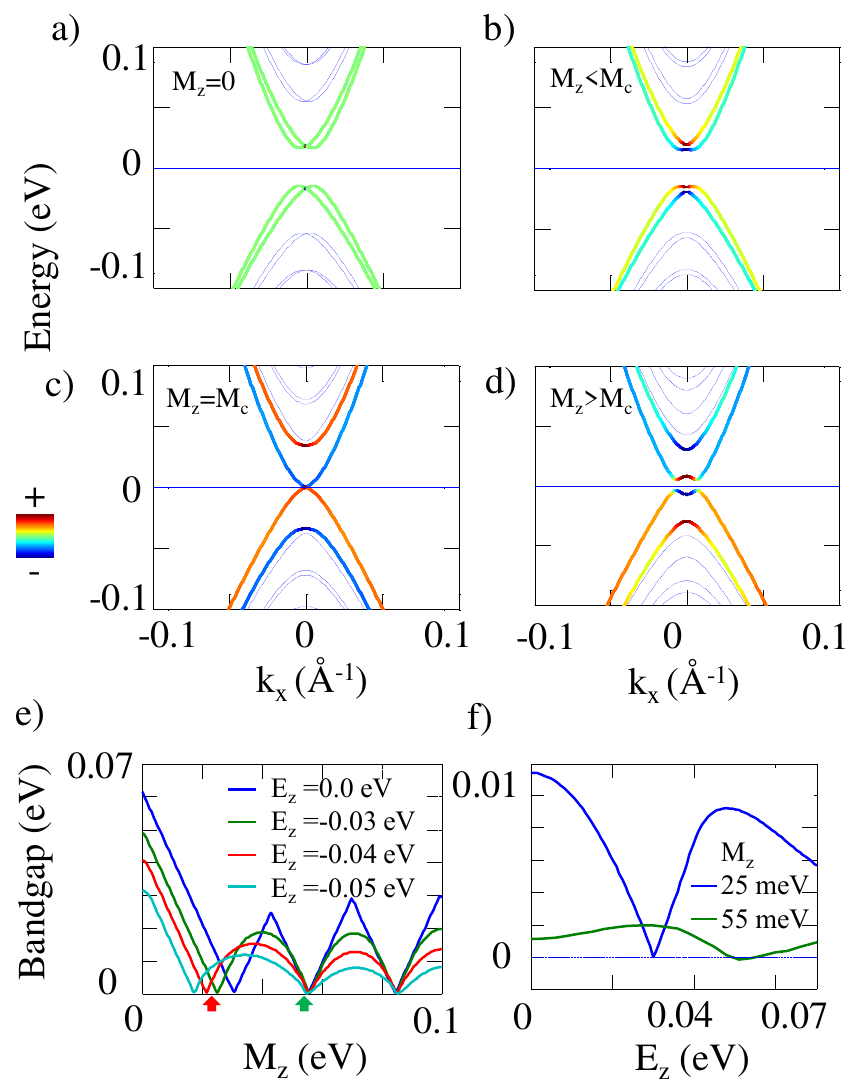}}
\caption{(Color online) The evolution of the band structure upon increasing $M_z$ with non-vanishing $E_z=-0.05$ eV: (a) $M_z=0$; (b) $0<M_z<M_c$; (c) $M_z=M_c$; (d) $M_z>M_c$, where the system turns to the $C=1$ QAH phase. The degree of out-of-the-plane spin polarization is denoted by color. (e) The non-linear band gap evolution as a function of $M_z$ with various electric field strength. Chern number $C$ (starting from 0) is added by one each time when the gap closes. The arrows shown here are the chosen exchange field strength to be compared in (f). (f) Electric field tunable QAH phases for two given exchange field strength. The gap closing points separate the $C=0$ and $C=1$ phases (blue curve), and the $C=1$ and $C=2$ phases (green curve), respectively.
} 
\end{figure}

\subsection{Electric field tunable QAHE}
One of the practical issues, based on the mechanism we have mentioned above, is the presence of certain substrates when preparing the thin films epitaxially in the experiments. It is equivalent to the presence of an effective electric field `$E_z$' along $z$ direction and this leads to a broken $z\rightarrow -z$ reflection symmetry for the thin films. Hence, we study systematically the effect of the electric field by adding a given linear potential term along $z$ direction, with potentials $\mp E_z$ on the top and bottom surfaces of the sample, respectively, in Eq.(\ref{eq2}) under various applied $M_z$. Note that the dispersion relation does not change if $E_z$ changes sign. 

We first consider the evolution of the energy spectrum near $\Gamma$ point as a function of $M_z$ with $E_z=-0.05$ eV, as shown in Figs. 5(a)-(d). In the absence of any  exchange field, the electric field simply introduces Rashba type interactions into the system and consequently each spin-degenerate band now splits with the originally band minimum shifted away from $\Gamma$ point, while the spin degeneracy still keeps intact at $\Gamma$ point [see Fig. 5(a)]\cite{crossover}. Assuming $N_l=2N$, it is worth noting that the wave function distribution of these lowest (highest) conduction (valence) bands around $\Gamma$ point is mainly from the contributions of $N$th and $N+1$th layers in the middle. This is in sharp contrast with the usual TI thin films ({\it i.e.,} $t_z > t_{zc}$) , in which the lowest conduction (highest valence) band has the largest weight from the top and the bottom layers. 

Upon turning on the exchange field, as one can see in Fig. 5(b), all the spin-degenerate points of Rashba-like bands at $\Gamma$ point open up gaps and the band gap of the system reduces to zero as $M_z$ reaches to a critical value $M_{c}$ [see Fig. 5(c)]. Further increasing $M_z$ results in a spin-polarized band inversion and reopens the band gap, leading again to the QAH phase [see Fig. 5(d)]. It is interesting to notice a few subtle differences from the case without $E_z$: 1) When $M_z=0$ the band gap is smaller due to the shift of the conduction band minimum; 2) at $M_z=M_{c}$, the energy dispersion is non-linear and the spin texture around $\Gamma$ point is relatively simple. Importantly, the above observations show that the critical exchange field strength is lowered and one can possibly achieve QAHE with high Chern number by tuning the electric field, as we explain next.

Fig. 5(e) illustrates the band gap evolution as a function of $M_z$ with a given electric field $E_z$ in an $N_l=20$ thin film. The band gap repeatedly closes and reopens, indicating that the system undergoes topological phase transitions several times up to the QAH phase with high Chern number ($C=3$ in our plot). Significantly, after considering several different values of $E_z$, we find that the critical exchange energies to achieve $C=1$ and $C=2$ phases, respectively, are less than the cases in the absence of the electric potential. To examine it more carefully, we take two representative initial phases in our system at $E_z=0$, as indicated by the arrows shown in Fig. 5(e): 1) a trivial $C=0$ phase with fixed $M_z$=0.025 eV and 2) a $C=1$ QAH phase with fixed  $M_z$=0.055 eV. Purposely, they are chosen just prior to QAH phases with $C=1$ and $C=2$ separately. We then compute the corresponding band gap as a function of $E_z$ for each of them. As one can see in Fig. 5(f), it is feasible to apply an external electric field to drive our focused system, namely, TlBi(S$_{1-\delta}$Se$_{\delta}$)$_{2}$ thin film from an originally Chern number $C$ QAH phase to another QAH phase with Chern number $C+1$. However, we would like to point out that this tuning approach is efficient to obtain QAH phase up to $C=2$. For getting higher Chern numbers, it might become unstable because several sub-bands would come into play around $E_F$ and the system may not maintain its insulating nature during the process.

\section{Conclusion}
In summary, we have presented our scenario to achieve QAH effect with field-tunable Chern number via a model study. Remarkably, the model can describe topological phase transition from a Z$_2$ trivial to an non-trivial insulating thin film for realistic materials such as TlBi(S$_{1-\delta}$Se$_{\delta}$)$_2$. By showing the band-structure evolution, spin-texture, and hence the spin-polarized band inversion, we clearly demonstrate the feasibility of our approach to tune the Chern numbers of the QAH phases through changing either the exchange field strength or the electric field strength in topologically {\it trivial} thin films near the Z$_2$ critical point (to TI phase). In particular, we stress that the necessary exchange field strength to exhibit high-Chern number QAH effect can be reduced further when pushing the system closer to the critical point and combining with the benefit from the substrate. 
Therefore, we hope that this paper could stimulate experimental works along this direction in the near future.

\begin{acknowledgments}
This research is supported by the National Research Foundation, Prime 
Minister's Office,  Singapore under its NRF fellowship 
(NRF Award No. NRF-NRFF2013-03).
W.F.T. acknowledges the support from MOST in Taiwan under Grant No.103-2112-M-110-008-MY3.
\end{acknowledgments}


\begin{thebibliography}{10}
\bibitem{warping}
When the warping effect is relatively small, such as in TlBi(S$_{1-\delta}$Se$_\delta$)$_2$, applying $M_z$ is the only effective way to make bands magnetically inverted; when the warping effect is large, such as in Bi$_2$Te$_3$, one may also apply in-plane magnetization along certain direction to induce QAH phase.
\end{thebibliography}
\end{document}